\documentclass[dvips,12pt]{article}
\usepackage{amsmath,amssymb,graphicx}
\newcommand\figcaption{\def\@captype{figure}\caption}
\newcommand\tabcaption{\def\@captype{table}\caption}

\def\vf{\varphi}
\def\ve{\varepsilon}
\def\Om{\Omega}

\def\dl{\delta}
\def\Dl{\Delta}
\def\la{\lambda}
\def\La{\Lambda}
\def\th{\theta}
\def\sg{\sigma}

\def\nn{\nonumber}
\def\diag{\mbox {diag}}
\begin{document}

\title{\LARGE\bf How to Measure the Cosmic Curvature}
\author{Ying-Qiu Gu\footnote {email: yqgu@fudan.edu.cn}}
\date{\small School of Mathematical Science, Fudan University, Shanghai 200433, China\\
\vspace{0.8cm} {\large M. Yu. Khlopov}\footnote{email:
Maxim.Khlopov@roma1.infn.it}\vspace{0.5cm} \\ Center for
Cosmoparticle physics, Cosmion, 125047, Moscow, Russia}

\maketitle

\begin{abstract}
The conventional method to determine the cosmic curvature is to
measure the total mass density $\Omega_{\rm tot}$. Unfortunately
the observational $\Omega_{\rm tot}$ is closely near the critical
value 1. The computation of this paper shows that $\Omega_{\rm
tot}\approx 1$ is an inevitable result for the young universe
independent of the spatial topology. So the mass density is not a
good criterion to determine the cosmic curvature. In this paper,
we derive a new criterion based on the galactic distribution with
respect to redshift $z$, which only depends on the cosmological
principle and geometry. The different type of spatial topology
will give different results, then the case can be definitely
determined.


\end{abstract}

\section{Introduction}
\setcounter{equation}{0}

Whether the space of our universe is open, flat or closed is one
of the most fundamental problems in cosmology. The solutions to
some other important problems such as the property of dark matter,
the evolution of the universe, are closely related to the spatial
topology of the universe. Almost all used methods to determine the
cosmic curvature are based on computing the total mass density
$\Om_{\rm tot}$. However, the most authoritative empirical data
show that the total mass density $\Om_{\rm tot}=1.02\pm
0.02$\cite{data1}-\cite{data4}, which is closely near the critical
density $1$, so we can not get a definite solution to this
fundamental problem.

Recent years, lots of attention have been attracted to the
resolution of this problem\cite{rul1}-\cite{flt8}. In \cite{rul3},
by analyzing the large-scale correlation function of red galaxies
from the Sloan Digital Sky Survey(SDSS) and combining with the
cosmic microwave background(CMB) acoustic scale, we arrive at
$\Om_K = -0.010 \pm 0.009$ under the assumption of $w=-1$. In
\cite{flt1}, by measured the large-scale power spectrum $P(k)$ for
the luminous red galaxies from SDSS, the authors improve the
evidence for spatial flatness, and sharpen the curvature
constraint to $\Om_{\rm tot} = 1.003\pm 0.010$.

In \cite{flt2,flt3}, using the full dataset of CMB and large scale
structure(LSS), and by Markov Chain Monte Carlo global fit, the
authors find the best fit value of $|\Om_K|<0.015$ for some dark
energy models and $|\Om_K|> 0.06$ are excluded. The $\La$CDM model
is consistent with all the data, and a flat universe is preferred.

From the calculation of the above papers, we find that the results
are strongly and nonlinearly correlated with the equation of state
$w(z)$. In \cite{oflt}, the authors solved $w(z)_F$ and $w(z)_d$,
two equations of state for the same model, respectively from the
Friedmann equation and the luminosity distance equation. The
numerical results show that the two functions $w(z)$ unstably
depend on $\Om_K$ and depend on $z$ towards opposite trends, which
lead to great errors for adequately large $z$. Similar conclusion
also confirmed in \cite{flt8}, where the authors find bounds on
cosmic curvature are less stringent if dark energy density is a
free function of cosmic time.

However, in our point of view, $\Om_{\rm tot}$ or $\Om_K$ may be
not the best criterion to determine the cosmic curvature. At
first, $\Om_{\rm tot}\approx 1$ is an inevitable results for the
young universe as shown bellow. Secondly, since the Friedmann
equation is a dynamic equation of $a(t)$, it is inadequate to
analyze this equation as algebraic equation by introducing
Hubble's parameter $H$. Thus looking for a more effective method
and criterion to determine the spatial type is necessary.

Hubble once realized to measure the spatial curvature by counting
galaxies\cite{count}, but the corresponding equation derived is
very complicated, which not only depends on $z$ but also on $H$
and $\Om_{\rm tot}$. This situation is caused by the calculation,
which is too direct to simplify the model, so that the idea was
not developed for practical application.

However, the geometrical problem may be more naturally and
effectively solved by geometrical method. In this paper we develop
the Hubble's idea by designing a simple reference function of the
galactic distribution in flat space, and then compare it with the
practical counting data. The result of comparison may be able to
give a definite answer to the fundamental question.

\section{Analysis and method}
\setcounter{equation}{0}

\subsection{Flatness problem of the space}

In mean sense, the universe is highly homogeneous and isotropic,
and the metric is described by Friedmann-Robertson-Walker(FRW)
metric. The corresponding line element is given by
\begin{equation}
ds^2=d\tau^2-a^2(\tau)\left(\frac {d
r^2}{1-Kr^2}+r^2(d\th^2+\sin^2\th d\phi^2)\right), \label{2.1}
\end{equation}
where $K=1,~ 0$ and $-1$ correspond to the closed, flat and open
universe respectively. Some literatures intended to replace the
constant $K$ by some continuous function $K(r)$ or $K(\tau)$. In
fact, such replacement violate the symmetry of the spacetime. The
cosmic radius of curvature is determined by $a(t)$ rather than by
$K$. The spatial topology of the homogeneous and isotropic
spacetime has only the three cases\cite{gem0}-\cite{gem3}, and the
metric must be equivalent to (\ref{2.1}). In this paper, we adopt
the conformal coordinate system, because in this form $a(t)$ can
be expressed as Taylor series near $a\to 0$. The corresponding
metric becomes
\begin{equation}
g_{\mu\nu}=a^2(t)\diag [1,-1,-f^2(r),-f^2(r)\sin^2\th],
\label{2.2}
\end{equation}
where $d\tau=a dt$, and
\begin{equation} f=\left \{ \begin{array}{ll}
  \sin r  & {\rm if} \quad  K=1,\\
   r   & {\rm if} \quad  K=0,\\
  \sinh r  & {\rm if} \quad  K=-1.
\end{array} \right. \label{2.3}
\end{equation}

The critical density $\rho_c$ and the Hubble's parameter $H$ are
defined by
\begin{eqnarray}
\rho_c=\frac 3{8\pi G}H^2\sim 8\times 10^{-30}({\rm g/cm}^3),\quad
H=\frac {a'(t)}{a^2} .\end{eqnarray} By Friedmann
equation\cite{count}, we have
\begin{equation}
\Om_K \equiv K \left(\frac a {a'}
\right)^2=\Om_m+\Om_\La-1=\Om_{\rm tot}-1,\label{2.5}
\end{equation}
where $\Om_m=\frac {\rho }{\rho_c},~ \Om_\La=\frac{\La}{3 H^2}$.
Theoretically, by (\ref{2.5}) we have the following judgment:
$\Om_{\rm tot}>1,~\Om_{\rm tot}=1$ and $\Om_{\rm tot}<1$
correspond to the closed, flat and open universe respectively. The
following analysis shows that, this judgement is not sharp for a
young universe.

The original equation of (\ref{2.5}) is the dynamic equation
\begin{equation}
a'^2=\frac 1 3\La a^4 + \frac{8\pi G}{3}\rho a^4-Ka^2.\label{2.6}
\end{equation}
For most matter models\cite{Lam}, such as nonlinear
spinors\cite{gu1,gu2, gu3} and Casimir effect, when $a\to 0$, we
have the main part of density as
\begin{equation}
\rho=\rho_0\left(\frac 1{a^3}-\frac {\sg}{a^4}\right), \label{2.7}
\end{equation}
where $\rho_0$ is a constant, and $|\sg|\ll a $ is a slowly
varying function of $a$, which can be treated as a constant for
the following calculation. For homogeneous scalar field models
such as quintessence and phantom field $\phi$ \cite{ess}, we can
also get the corresponding $\rho(a)$ via transforming $\frac
{d\phi}{dt}=\frac{d \phi}{da} a'$ and then solving the
differential equation. However, we need not to solve the specific
model for the present purpose, because we only need the assumption
that the cosmic equation of state $\rho(a)$ exists and has
singularity not worse than (\ref{2.7}). There are a lot of works
on constructing complicated equation of state $w(z)$ and fitting
the empirical data. It seems unnecessary, because we can hardly
find out the effective $w(z)$ \cite{oflt}.

We take (\ref{2.7}) with constant $\sg$ as model to show the
properties of $a(t)$ and $\Om_K$ for small $t$. We take $R=\frac
{4\pi G} 3 \rho_0=1$ as length unit, which is the mean scale
factor. Substituting (\ref{2.7}) into (\ref{2.6}) we get
\begin{equation}
a'^2=2(a-\sg)-Ka^2+{\la} a^4, \label{2.8}
\end{equation}
where $\la=\frac 1 3 \La R^2$ is dimensionless constant. Choosing
suitable starting time $t_0=0$, for small $t$ the solution of
(\ref{2.8}) can be expressed by
\begin{eqnarray} a (t)&=&a_0 +\frac 1
2(1-K a_0 +2\la a_0^3)t^2+\label{2.9} \\  &~& \left(-\frac K{24}+
\left( \la\sg+\frac {K^2}{24} \right) a_0 -\frac {3\la} 4
a_0^2+\frac {\la K} 6 a_0^3 \right) t^4+O(t^6),\nn
\end{eqnarray}
where $0 \le |a_0|\sim |\sg|\ll 1$ is the root of $2(a-\sg)-
Ka^2+\la a^4=0$ near zero. For the Big Bang model we have $a_0=0$.
(\ref{2.9}) shows how the parameters $K,~\sg,~\la$ and $a_0$
influence the scale factor.

\begin{figure}
\centering
\includegraphics[width=12cm]{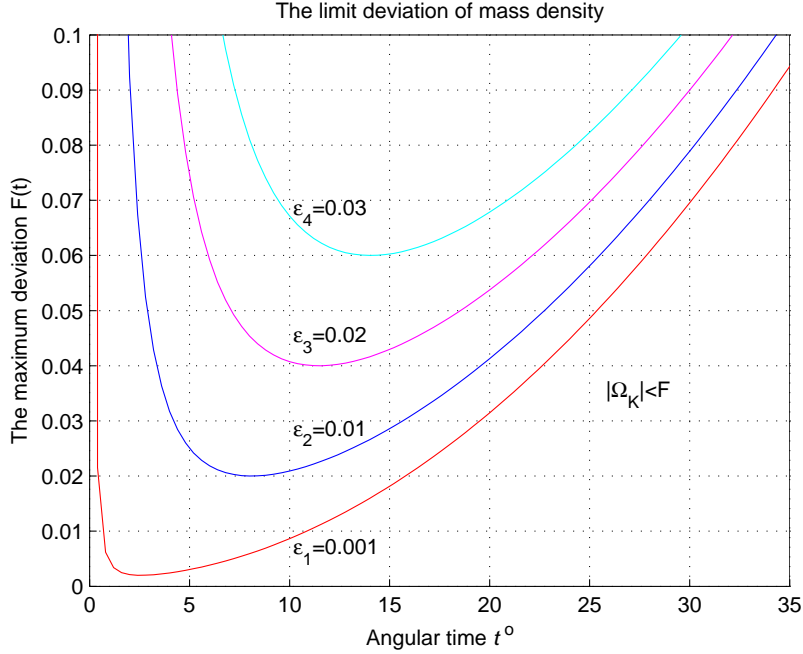}
\caption{The Maximum Deviation of Mass Desity Caused by Space
Type} \label{fig1}
\end{figure}

As computed in \cite{gu1,gu2}, we have the following conclusions.
(I) Since $\la$ has effects only on the cosmic scale, we have
$|\la|< 1$. (II) The present comving time $t_a\sim
18^\circ=0.314({\mbox{rad}})$. In this paper, we are only
concerned with the behavior of the universe after the galaxies
formed, so we have the estimation for maximum of $|\Om_K|$ as
follows
\begin{eqnarray} |\Om_K|\le\frac
{a(t)^2}{a'(t)^2}\dot=\left(\frac \ve t + \frac t 2\right)^2\equiv
F(t),\quad \ve\equiv \frac {a_0}{1-Ka_0+2\la a_0^3}\to 0.
\label{2.10}
\end{eqnarray}
The typical values are displayed in Fig.(\ref{fig1}), which
visually tells us $\Om_K$ is not a good criterion to determine the
cosmic curvature. If $a_0 \le 0$, then $\Om_K(t)=0$ has solution
$t_{{\mbox{singular}}}\ge 0$, which inevitably leads to the so
called ``fine-tuning mass density'', thus the dynamic process can
not be understood with static equation.

\subsection{Galactic distribution and reference function}

Now we design the new method to determine the cosmic curvature
$K$, which only depends on the redshift distribution of galaxies
at a few selected point $z_k$. The cosmological redshift
\begin{equation}z=\frac{\la-\la_0}{\la_0}=\frac
{a(t)}{a(t_0)}-1,\label{2.11} \end{equation} where $t=t_0+r$ is
the present time, and $t_0$ is the time when the photon was
emitted, $r$ is the radial coordinate of the light source. Denote
the number of all galaxies in a ball with radius coordinate $r$ by
$N(r)$, we consider $N(r)$ in the domain $[r-\frac 1 2 \Dl r,
r+\frac 1 2 \Dl r]$, or correspondingly in $[z-\frac 1 2 \dl z,
z+\frac 1 2 \dl z]$. The increment of redshift is given by
\begin{equation} \dl z\equiv \Dl \frac
{a(t)}{a(t_0)}=\left|\frac{a(t)a'(t_0)}{a^2(t_0)}\right|\Dl
t=(1+z)\left|\frac{a'(t_0)}{a(t_0)}\right|\Dl r,\label{2.12}
\end{equation} where we used $\Dl t=\Dl r$ for photons in the metric (\ref{2.2}). Then the
galaxies in the domain $[r-\frac 1 2 \Dl r, r+\frac 1 2 \Dl r]$
can be counted by
\begin{equation}
\dl N=4\pi \rho_g f^2(r) \Dl r={4\pi \rho_g
f^2}\left|\frac{a(t_0)}{a'(t_0)}\right|\frac{\dl
z}{1+z},\label{2.13}
\end{equation} where $\rho_g$ is the number density of galaxies in
comoving volume $d V=f^2\sin\th dr d\th d\phi$, which is a
constant independent of $t$. By (\ref{2.13}), we get the following
rigorous equation
\begin{equation}(1+z)\frac{d N}{d
z}\left|\frac{a'(t_0)r}{a(t_0)}\right|=4\pi\rho_g
f^2r.\label{2.14}
\end{equation}

By (\ref{2.9}) and (\ref{2.11}), we can solve $r=r(t,z)$, then get
the following function
\begin{eqnarray}
\vf\equiv \left|\frac{a'(t_0)r}{a(t_0)}\right|. \label{2.15}
\end{eqnarray}
Generally speaking, $\vf=\vf(t,z)$ is a complicated function.
However for a small redshift, by the empirical Hubble's law, we
have
\begin{eqnarray}\vf=\left|\frac{a'(t_0)r}{a(t_0)}\right|
=\left|\frac{a'(t_0)}{a(t_0)^2} \right| D_r =H(t_0) D_r\dot =z,
\label{2.16} \end{eqnarray} where $D_r=a(t_0) r$ is the distance
corresponding to radial coordinate $r$ at time $t_0$. For flat
space, the number of galaxies in a ball with radius $r$ is given
by
\begin{equation}
N(r)=\frac{4\pi}3\rho_g r^3=\frac{4\pi}3\rho_g
f^2r.\label{2.17}\end{equation} Substituting (\ref{2.16}) and
(\ref{2.17}) into (\ref{2.14}), we get the distributive equation
for flat space
\begin{equation}
\frac{dN}{Ndz}\dot = \frac{3}{z(1+z)}.\label{2.18}\end{equation}
The solution of (\ref{2.18}) is given by
\begin{equation}
N(z)\dot =  N_1 \left(\frac z
{1+z}\right)^3.\label{2.19}\end{equation} By the precision of
Hubble's law, we learn (\ref{2.19}) is a good approximation for
galaxy count at lest within the range $z<1$.

However, for the flat space $K=0$, we can derive more accurate
$N(z)$ for large range $z$ as follows. By (\ref{2.9}), we always
have solution to high accuracy
\begin{equation}
a(t)\dot= \frac 1 2t^2 , \qquad ({\mbox{for}}~~ 5^\circ\le t\le
60^\circ), \label{2.20}
\end{equation}
$t\ge5^\circ$ is because only the evolution after galaxies formed
is concerned here, then we can omit the influence of $a_0$. The
condition $t\le 60^\circ\sim 1$(rad) is derived by omitting the
higher order term in the series (\ref{2.9}).

Substituting (\ref{2.20}) into (\ref{2.11}), we get
\begin{equation} z=\frac {(t_0+r)^2-t_0^2}{t_0^2}=\frac {r}{t_0}\left(2+\frac
{r}{t_0}\right),\label{2.21}\end{equation} or equivalently
\begin{equation} \frac {r}{t_0}=\sqrt{1+z}-1=\frac
z{1+\sqrt{1+z}},\label{2.22}\end{equation} so we have
\begin{equation} \vf=\frac {a'(t_0)}{a(t_0)}r=\frac {2r}
{t_0}=\frac {2z}{1+\sqrt{1+z}}.\label{2.23}\end{equation}
Substituting (\ref{2.23}) and (\ref{2.17}) into (\ref{2.14}), we
get the distributive equation for flat space to high accuracy
\begin{equation}
\frac{dN}{Ndz}=
\frac{3(1+\sqrt{1+z})}{2z(1+z)}.\label{2.24}\end{equation} The
solution of (\ref{2.24}) is given by
\begin{equation}
N(z)= N_0 \left(\frac z
{1+z+\sqrt{1+z}}\right)^3.\label{2.25}\end{equation}

\begin{figure}
\centering
\includegraphics[width=12cm]{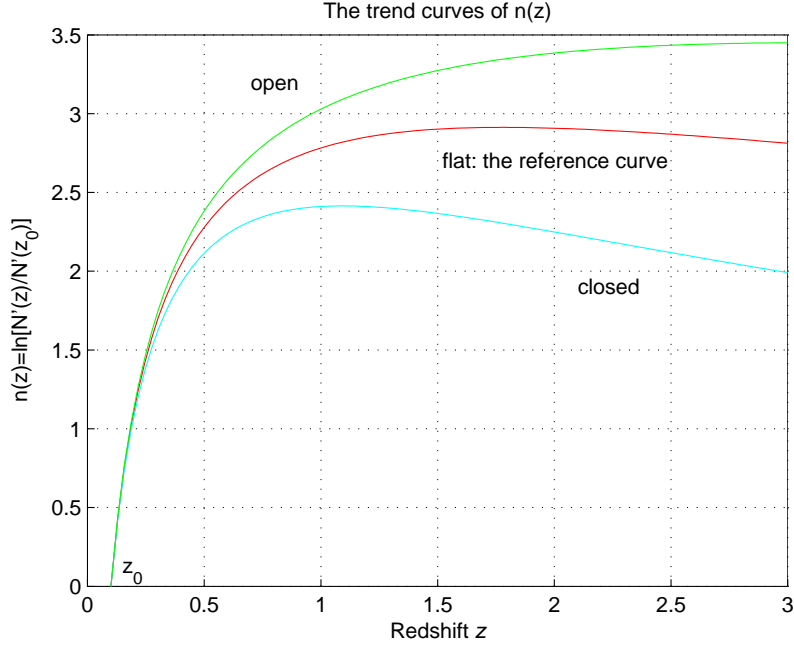}
\caption{The Distributive Functions of Counting Galaxies}
\label{fig2}
\end{figure}

Define the reference function by
\begin{equation}
n(z)\equiv \ln\left(\frac{N'(z)}{N'(z_0)}\right)=
2\ln(\sqrt{1+z}-1)-\frac 5 2 \ln(1+z)-n_0,
\label{2.26}\end{equation} where $z_0$ is the first redsheft point
measured, from which we count galaxies,
\begin{eqnarray}n_0=2\ln(\sqrt{1+z_0}-1)-\frac 5 2 \ln(1+z_0)\end{eqnarray} is a constant.
The function $n(z)$ is the criterion which we are looking for,
because for all $z>z_0$, we definitely have
\begin{equation}
n(z)_{\rm open}>n(z)_{\rm flat}=2\ln(\sqrt{1+z}-1)-\frac 5 2
\ln(1+z)-n_0>n(z)_{\rm closed},\label{2.28}\end{equation} due to
the direct influence of $f(r)$ on $N(z)$. The trend curves of
$n(z)$ are displayed in Fig.(\ref{fig2}). The practical counting
data $n(z)$ should also be a smooth function of $z$ due to the law
of large number. By this method, we will compare the trends of two
smooth functions rather than two data, so there should be less
ambiguous consequence caused by experiment and model errors.

\section{Discussion and conclusion}
\setcounter{equation}{0}

From the above computation and analysis, we can get the following
results.

(C1). From Fig(\ref{fig1}) we learn that the present mass energy
$\Om_K$ corresponding to the spatial type is generally small in
value for a young universe, which is not sharp to give a definite
answer for the spatial type.

(C2). The space type can be determined by counting the galactic
distribution with respect to the redshift $z$, which is a smooth
function of $z$ and can be sharply compared with the theoretical
reference function (\ref{2.26}).

(C3). The cosmic equation of state $\rho(a)$ is a useful concept
for cosmology, because $\rho$ is a superposable intensity with
clear physical meanings, and it can be easily derived for concrete
matter models via action principle, which seems to be more
effective than $w(z)$.

\section*{Acknowledgments}

The author Y. Q. Gu is grateful to his supervisor Prof. Ta-Tsien
Li for his guidance and help.


\begin{thebibliography}{99}
\bibitem{data1} A. G. Riess {\em et al.} (Supernova Search Team), Astron. J. 116,
1009 (1998), \\ astro-ph/9805201.
\bibitem{data2} S. Perlmutter {\em et al.} (Supernova Cosmology Project), Astrophys. J. 517, 565 (1999),
astro-ph/9812133.
\bibitem{data3} N. Spergel {\em et al.} (WMAP), Astrophys. J. Suppl. 148, 175 (2003), \\ astro-ph/0302209.
\bibitem{data4} M. Tegmark {\em et al.} (SDSS), Phys. Rev. D69, 103501 (2004), astro-ph/0310723.
\bibitem{rul1} T. Broadhurst, A. H. Jaffe, {\em USING THE COMOVING MAXIMUM OF THE GALAXY POWER SPECTRUM TO
MEASURE COSMOLOGICAL CURVATURE}, astro-ph/9904348v1
\bibitem{rul2} B. F. Roukema, G. A. Mamon, {\em Tangential Large Scale Structure as a Standard Ruler:
Curvature Parameters from Quasars}, A \& A, 358, 395(2000),\\
astro-ph/9911413
\bibitem{rul3} D. J. Eisenstein, {\em et al, DETECTION OF THE BARYON ACOUSTIC PEAK IN THE LARGE-SCALE
CORRELATION FUNCTION OF SDSS LUMINOUS RED GALAXIES},
astro-ph/0501171
\bibitem{flt1} M. Tegmark {\em et al. Cosmological Constraints from the SDSS Luminous Red Galaxies} astro-ph/0608632;
\bibitem{flt2} Gong-Bo Zhao, Jun-Qing Xia, {\em et al.} {\em Probing for dynamics of dark energy and curvature of universe with
latest cosmological observations}, astro-ph/0612728
\bibitem{flt3} E. L. Wright, {\em Constraints on Dark Energy from Supernovae, $\gamma$-ray bursts, Acoustic Oscillations, Nucleosynthesis and Large
Scale Structure and the Hubble}, \\ astro-ph/0701584v3
\bibitem{flt4} B. Mota, {\em et al,  Constraints on the Detectability of Cosmic
Topology from Observational Uncertainties}, gr-qc/0308063
\bibitem{flt5} K. Ichikawa, T. Takahashi, {\em Dark Energy Parametrizations
and the Curvature of the Universe}, astro-ph/0612739v2
\bibitem{flt7} Y. G. Gong, A. Z. Wang, {\em Reconstruction of the deceleration parameter and the equation of state of dark energy}, astro-ph/0612196v3
\bibitem{oflt} C. Clarkson, M. Cortes, B. Bassett, {\em Dynamical Dark Energy or Simply Cosmic Curvature?}, astro-ph/0702670
\bibitem{flt8} Y. Wang, P. Mukherjee, {\em Observational Constraints on Dark Energy and Cosmic Curvature}, astro-ph/0703780
\bibitem{count} M. S. Longair, {\em Galaxy Formation}(Ch8, Ch17.2), Springer, (2001)
\bibitem{gem0} S. L. Weinberg, {\em Gravitation and Cosmology}(Ch.13, Ch.14), Wiley, New York, 1972
\bibitem{gem1} M. Lachi\`eze-Rey, J.-P. Luminet, {\em COSMIC TOPOLOGY}, gr-qc/9605010
\bibitem{gem2} J.-P. Luminet,  B. F. Roukema, {\em TOPOLOGY OF THE UNIVERSE: THEORY AND OBSERVATION}, astro-ph/9901364
\bibitem{gem3} J. Weeks, R. Lehoucq, J.-P. Uzan, {\em Detecting Topology in a Nearly Flat Spherical Universe}, astro-ph/0209389
\bibitem{Lam} M. Szydlowski, A. Kurek, A. Krawiec, {\em Top ten accelerating cosmological models}, Phys. Lett. B642(2006) 171-178, astro-ph/0604327
\bibitem{gu1} Y. Q. Gu, {\em A Cosmological Model with Dark Spinor
Source}, IJMPA Vol.22, No.25 (2007) 4667-4678, gr-qc/0610147
\bibitem{gu2} Y. Q. Gu, {\em Accelerating Expansion of the Universe with Nonlinear Spinors}, \\ gr-qc/0612176
\bibitem{gu3} Y. Q. Gu, {\em Thermodynamics of Ideal Gas in Cosmology},
arXiv:0708.2962
\bibitem{ess} Z. K. Guo, Y. S. Piao, X. M. Zhang, Y. Z. Zhang, {\em Two-Field Quintom Models in the $w-w'$ Plane}, astro-ph/0608165

\end{thebibliography}
\end{document}